# DETERMINATION OF FRIENDSHIP INTENSITY BETWEEN ONLINE SOCIAL NETWORK USERS BASED ON THEIR INTERACTION

*Sanja Krakan, Luka Humski, Zoran Skočir*

Online social networks (OSN) are one of the most popular forms of modern communication and among the best known is Facebook. Information about the connection between users on the OSN is often very scarce. It's only known if users are connected, while the intensity of the connection is unknown. The aim of the research described was to determine and quantify friendship intensity between OSN users based on analysis of their interaction. We built a mathematical model, which uses: supervised machine learning algorithm Random Forest, experimentally determined importance of communication parameters and coefficients for every interaction parameter based on answers of research conducted through a survey. Taking user opinion into consideration while designing a model for calculation of friendship intensity is a novel approach in opposition to previous researches from literature. Accuracy of the proposed model was verified on the example of determining a better friend in the offered pair.

**Određivanje intenziteta odnosa prijateljstva među korisnicima na društvenoj mreži temeljem njihova mrežnog međudjelovanja**

Izvorni znanstveni članak

Društvene mreže jedan su od najpopularnijih načina moderne komunikacije, a Facebook je najpopularnija društvena mreža. Informacije o povezanosti korisnika na društvenoj mreži često su vrlo šture. Zna se tek je li netko s nekim povezan, ali intenzitet te povezanosti nije poznat. Cilj našeg istraživanja jest odrediti i kvantificirati intenzitet prijateljstva među korisnicima društvene mreže na temelju analize njihova međudjelovanja na društvenoj mreži. Osmislili smo matematički model koji koristi: algoritam nadziranog strojnog učenja nasumične šume, eksperimentalno određene koeficijente značajnosti za pojedine načine komunikacije te koeficijente značajnosti za pojedine načine komunikacije izračunate na temelju odgovora korisnika u provedenoj anketi. Uzimanje u obzir mišljenja korisnika o važnosti pojedinih načina komunikacije na društvenoj mreži, pri izgradnji modela za određivanje intenziteta odnosa, novi je pristup u odnosu na ostale radove koje smo imali priliku pročitati. Točnost rada predloženog modela ispitana je na problemu određivanja boljeg prijatelja u paru prijatelja.

**Ključne riječi:** društvene mreže; Facebook; matematički model; interakcija korisnika; nasumične šume; nadzirano učenje; strojno učenje; težina veze

## 1 Introduction

Around 40% of today's world's population has an Internet connection. In 1995 less than 1% of the world's population were Internet users. Their number has increased tenfold from 1999 until 2013. First billion users were reached in 2005, the second billion in 2010 and the third billion in 2014. The Internet currently counts 3,4 billion users and growing rapidly [1]. It makes everyday life easier by allowing simplified search of desired information, performing different tasks and enabling communication between its users.

One of the most popular forms of modern interaction and communication are online social networks (OSN). OSNs are online communities of people that allow individuals to create a public or a semi-public profile, enabling them to form a list of other users with whom they wish to be connected to. Users have the possibility to browse through each other's profiles and interact by employing various technologies available on these networks [2].

Real life relationships can be classified in various terms based on their quality and intensity, nevertheless, all connections on OSN regardless of their "real life" relationship status are summarized into one type – friendship. Furthermore, users of online social networks often have lists of connections that also contain unknown users such as e.g. public figures who fall under the same "friendship" category as "real life" friends and acquaintances. Due to this lack of relationship differentiation based on its quality and intensity as discussed above, online social networks often have a difficult task deciding which information and recommendation to display to their users, how to provide a better and a more interesting service, and to whom to promote certain products. This sets an interesting research question, **is it possible to determine friendship intensity of online social network users based on their interaction online?**

Different communication parameters are used to achieve user communication on OSNs. These communication parameters, as a form of interaction among users, are private messages, comments, likes, photo tags etc.

For research described in this paper, insight into the importance of every communication parameter respectively from the perspective of users was gained from research conducted through a survey. Anonymous data from Facebook collected in the previous research was used in order to design a mathematical model to determine friendship intensity between Facebook users based on their interaction.

The main hypothesis of this paper is: **Greater the weight of friendship calculated for two observed individuals according to the proposed model, the stronger the intensity between them exists in the real life**. It is believed that the stronger link establishes a more trusted relationship between friends, family and like-minded people in general.

The paper is organized as follows: section 2 describes related work; section 3 introduces a model for calculating friendship weight and describes a survey carried out; section 4 presents the results of the research; section 5 provides discussion of the whole research and the results and section 6 gives a conclusion and elaborates ideas for future research.

## 2 Related work and motivation

The past few years note a rise in the number of papers and researches that aim to develop a model to determine the friendship intensity or the link weight between two users based on their interaction. It is assumed that the interaction between strongly linked users will be more frequent than between weak linked users, in other words, those who are not mutually close.

Today's researches are mutually interlaced and similar parameters of interaction are being used on various OSNs. A common characteristic is collecting two types of information: data from online social networks about users, their actions and interaction, and users' assessment of the observed relationship that is considered as ground truth [3][4][5][6][7][8][9][10].

There are three approaches. Development of the mathematical model and the experimental determination of the coefficients for multiplication [6][8][9][10], the use of the supervised machines learning algorithms [11][12][13][14] or a combination of both [15]. The aim is unique, detection of the link between users and its strength.

The novelty in this type of research is the use of the semantics bond with the basic use of communication parameters, i.e. mention of certain persons in posts [16][17].

The need for knowing the intensity of relations between users appears in different areas. The primary application is to use a model in order to improve the service of OSN [9]. Telecoms are trying to detect possible churners (users that are likely to change network) by analyzing social network where a stronger link or tie means a greater influence between users [18][19][20]. Usually information for building that kind of social network is fetched from call detail records (CDR). Enterprises would prefer to see a correlation between tie strength of their employees on online social networks and a level of cooperation and communication between them [21][22]. That kind of social network is built by a process of analyzing communication of employees through different corporation communication channels. In order to promote their brand through online marketing, enterprises are also analyzing OSNs to detect users with a greater impact on their online friends [6][22]. The application of "a stronger tie or friendship intensity equals more influence" model is evident in the use of recommendation system because it is expected that the user will have more trust in their real life close friends [12][22]. It is also worth mentioning that this type of model can be used in order to suppress crime, frauds and terrorism, which is an ongoing issue today [6][7]. It is highly probable that fraudsters, terrorists and criminals interact with users of similar intentions on online social networks. Therefore, it would be of a great value to identify more trusted, close and in general like-minded friends of an observed user in order to perform targeted recommendations of products and services or to expose a network of criminals, terrorists or fraudsters.

The essential benefit of statistical analysis of social networks is a better interpretation of users' interaction. The ultimate aim of this research is basically equal to previous researches, to determine the importance of parameters of interaction, but the initial approach is somewhat different. While other researches mostly attempted to achieve the distribution of parameters by analyzing communication, this research employs a survey as a starting point for the development of a precise mathematical model. The purpose of the survey is to obtain a better view on the importance of interaction parameters from the perspective of the originators of interaction i.e. OSN users.

## 3 Methodology

### 3.1 Survey

The survey by questionnaire was conducted with the purpose of obtaining an enhanced insight into the usage of communication parameters on Facebook from the users' perspective. The aim was to facilitate the decision which parameters to use and how to use them while developing the mathematical model. Participants had to answer 13 mandatory questions. It was conducted in English in order to get a more global insight.

Participants were asked to give their opinion on the interrelationship between the usage of interaction parameters (likes, comments, chat, tags in statuses and photos) and the real life friendship intensity. The aim was to determine if the higher value of parameters, i.e. more communication or more likes, means greater closeness in real life.

Besides Messenger[1] mobile application stores offer a variety of other communication applications (Viber, WhatsApp etc.) so insight into their use in relation to "real life" relationship status is considered relevant. The goal was to determine the Messenger communication proportion of the total communication through such applications. Participants were also asked about the use of *Close Friends* list on Facebook and if they add only "real life" close friends. The objective was to distinguish parameters of interest as opposed to those in need of elimination for the sake of complexity reduction.

### 3.2 Used data, cleaning and preparation for modeling

In this research we use the data set collected in previous researches [7][8]. That data set contains anonymized information about the interaction of the respondents who are Facebook users (ego-users) and all their network friends [7][8]. Respondents had to choose a better friend in the offered pair, classify friends into subgroups and create a list of up to 10 best friends. Ego-user's (subjective) assessment of the observed relationship is considered as ground truth.

In further modeling we used only part of the mentioned data set. We used the following data tables: a data table with 231 055 records containing all ego-users, their friends and 15 interaction parameters for each pair *(ego-user, friend)*; a data table with 23 936 records containing the classification of random friends into four subgroups: *best friends, friends, acquaintances and*

---
[1] Messenger is the official Facebook application, which lets Facebook users to have text conversations with people from their friend list.

*unallocated* and a data table with 16 991 records containing ego-user, a random pair of his friends and ego-user's selection of the better friend in a pair.

Cleaning and preparation of the data is one of the most important parts of statistical analysis, which often requires more time dedication than data analysis. There is a range of techniques, implemented in the R[2] environment, that allow the creation of scripts for cleaning of the data that contain a wide range of errors and inconsistencies [23].

Using R scripts data set was cleaned of all missing values. For each unique ego-user, interaction towards their friends was summed up by a type of interaction. The aim was to prefilter and eliminate unimportant data, and select the most relevant instances. We wanted to detect users which use Facebook passively and exclude them from the research. It was very difficult to accurately determine the set of threshold values for the sum of interaction bellow which users will be marked as passive users and eliminated. Serval times the number of exchanged messages was confirmed as one of the most important interaction parameter and, consequently, our main orientation point. By observing total users' sum of interaction, divided by a type of interaction, we decided to use 10[th] percentile as a threshold for separation of passive and active users. Therefore, it was decided that all users who have less than 2100 exchanged messages with all of their friends and simultaneously all other interaction parameters less than 3, will be eliminated from the data set. After that cleaning data set process, the table with all pairs *(ego-user, friend)* was reduced to 230 397 records. From the table containing the classification of friends into subgroups, we eliminated friends placed in the subgroup labeled *unallocated,* in order to use a more exact data set. The table was reduced to 23 407 records. Combination of these two tables is used for performance of the supervised machine learning algorithms. Further, from the data table containing users decision about the better friend in pair, we removed all pairs for which user couldn't decide on a better friend, since it's evident that with the use of the mathematical model, most of the time exclusively one or the other friend will be chosen, in other words, the decision will fall on the one that has the greater weight calculated. This table was reduced to 13 235 records and it's used in the process of the verification of the model we propose in this paper.

Additionally, all numerical parameters of interaction have been normalized according to [24]. Parameters are scaled in the range [0,1] separately using formula (1).

$$x_{norm,i} = \frac{x_i - x_{min}}{x_{max} - x_{min}} \qquad (1)$$

$x_i$ is *i*-th value of the observed parameter, $x_{min}$ is the minimal value of that parameter, $x_{max}$ the maximal value and $x_{norm,i}$ normalized *i*-th value. Normalization techniques are used to transform various parameters of a large range of values to a lower range in order to facilitate processing [24].

---

[2] R is a language and environment for statistical computing and graphics.

### 3.3 Development of the mathematical model

Previously prepared data set with the data about interaction, as predictor variables, and the classification, as ground truth, was used in the process of the development of the mathematical model. Linear Regression and Random Forest (supervised machine learning algorithms) were performed on this data set.

Mathematical models were built and verified gradually using several interaction parameters.

First proposed model uses just the parameter importance obtained with the Linear Regression and it's given with (2). In the Linear Regression, importance of each parameter can be represented with the absolute *t value* [25]. The higher the *t value* is, greater the importance of individual parameter of the interaction is.

$$friendship\_weight = \sum_{i=0}^{n} t\_value_i * x_{norm,i} \qquad (2)$$

The second proposed model uses just parameter importance obtained with the Random Forest algorithm and it's given with (3). In the Random Forest importance of each parameter can be represented with the *MeanDecreaseAccuracy* [25][26]. The higher the value of *MeanDecreaseAccuracy* is, greater the importance of individual parameter of interaction is.

$$friendship\_weight = \sum_{i}^{n} MeanDecreaseAccuracy_i * x_{norm,i} \qquad (3)$$

Verification has shown that the model (3), which uses the Random Forest algorithm, gives better results than the model (2), which uses the Linear Regression (table 9 and 10). Therefore, it was decided that the model (3) would be used as a starting point for further improvements. Compared to the model (2), the model (3) gives approximately 2% better results and it's decided that the parameter importance can be represented by the value of *MeanDecreaseAccuracy* variable.

Next proposed model (5), in addition to the *MeanDecreaseAccuracy,* uses the distribution of interaction parameters $p_i$, which is calculated according to the answers from the survey using (4).

$$p_i = \frac{b}{u} * \frac{1}{v} \qquad (4)$$

Number of users who said that the observed parameter *i* is the most important in the communication towards real life friends is described with $b$; $u$ is the total number of answers and $v$ number of parameters that use the same type of interaction. Same type of interaction means that we summarized in one all comments or all likes, no matter which object is commented or liked, e.g. comments on photos and wall posts together.

$$friendship\_weight = \sum_{i}^{n} MeanDecreaseAccuracy_i * p_i * x_{norm,i} \qquad (5)$$

In order to give more importance to certain parameters, coefficient $k_i$ was experimentally determined. The coefficient $k_i$ is from the interval [1, 100] and it was calculated for each of the parameters of interaction while keeping all the other parameters constant. The change in accuracy of the model was observed.

The final proposed model for the determination of friendship intensity between online social network users based on their interaction uses the *MeanDecreaseAccuracy*, the distribution of interaction parameters $p_i$, coefficients $k_i$ and it's given by (6).

$$friendship\_weight = \sum_{i=1}^{n} MeanDecreaseAccuracy_i * p_i * k_i * x_{norm,i} \quad (6)$$

### 3.3.1 Example of using mathematical model

Examples of the calculation using the final model will be given in this sub-section, as well as the verification example demonstrated on the random pair of friends. Used normalized numerical interaction parameters for two pairs *(ego-user, friend)*, pairs *(A,B)* and *(A,C)* are shown in table 1. Values of *MeanDecreaseAccuracy* are given in the table 5, $p_i$ in the table 6 and $k_i$ in the table 8.

**Table 1** Example of normalized value of interaction parameters between ego-user A and his friends B and C

|  | Value of i-th interaction towards friend B | Value of i-th interaction towards friend C |
|---|---|---|
| Observed friend's *comments on a wall post* of ego-user A | 0 | 0 |
| Observed friend's *messages on the wall* of ego-user A | 0 | 0 |
| Ego-user A and observed friend are *tagged together in a post* | 0 | 0 |
| *Mutual photo* published by ego-user A | 0,016260163 | 0 |
| *Mutual photo* published by a user that is not ego-user A or his observed friend | 0,006097561 | 0,006097561 |
| Observed friend comments on a photo of ego-user A | 0,004405286 | 0 |
| *Messages* exchanged between ego-user A and his observed friend | 0,0336021542 | 0,0134639900 |
| Ego-user A used *close friend* option for observed friend | 0 | 0 |

According to the model (6) and the values from the table 1, friendship weight can be calculated from ego-user A towards his friends B and C. Data for the verification of accuracy of the model on an example is shown in the table 2.

Ego-user's (subjective) assessment of the observed relationship is considered as ground truth. Ego-user said that the first friend in the offered pair (B) is better, i.e. closer friend in real life.

According to calculation with (6):
$friendship\_weight(A, B) = 20,49882$
$friendship\_weight(A, C) = 8,196606$.
$friendship\_weight(A, B) > friendship\_weightht(A, C)$,
model prediction is accurate, i.e. equal to ego-user's answer.

**Table 2** Verification of the model's (6) accuracy on given example

| Ego-user's A statement | 1 |
|---|---|
| Link weight from A to B | 20,49882 |
| Link weight from A to C | 8,196606 |
| Model prediction | 1 |
| Model prediction = user statement | True |

## 4 Results

### 4.1 Survey

In this section results of the conducted survey will be presented. Participants had to answer 13 mandatory questions. A survey was conducted in English in order to get a more global insight. Of all 144 respondents, 94,4% use Facebook every day, 4,2% use it 4-5 times a week, and 1,4% only 2-3 times a week. None of the participants responded positively to the statement that they use Facebook several times a month or that they don't use it often.

Questions that were carrying the most information will be elaborated more closely. Most respondents said that they use Facebook to browse through friends' posts and communicate through Messenger.

Facebook users can choose to subdivide friends from their Facebook friends list by adding chosen friends to a sub list called *Close Friends*. A list of close friends is a smart list and a user receives a notification when his friends in this list refresh their status or perform any other action. Users were asked whether and how do they use the close friends list. Replies are shown on the graph 1.

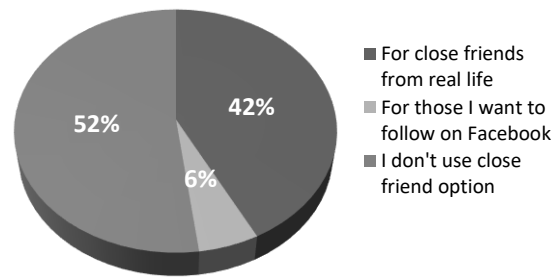

**Graph 1** Relation of Close Friends list on Facebook and real life close friends

One of the available parameters of interaction is a number of mutual friends, which indicates how much friends lists between two Facebook users overlap. Participants were asked whether they think a larger number of mutual friends means greater closeness in the real life. Only 16% of users responded affirmatively, while 84% believe that the number of mutual friends on Facebook and the closeness in real life are not correlated.

Participants were asked: "In your opinion, does a number of likes on Facebook friends posts indicate that you are most likely friends in real life? For example, if you liked 50 posts from a certain Facebook friend, does it indicate that this person is necessarily a better friend in the real life than someone whose posts you liked none or a few times?" Respondents had to choose an answer on the scale of 1 to 4. Statement number 1 means negative indication, i.e. users like posts from the best friends and acquaintances equally depending on the content of a post. Statement number 4 means positive indication, users mostly like real life friends' posts. Answers number 2 and 3 mean that the user doesn't agree completely with either statement, but leaning more to either statement number 1 or 4. Replies are shown on the graph 2.

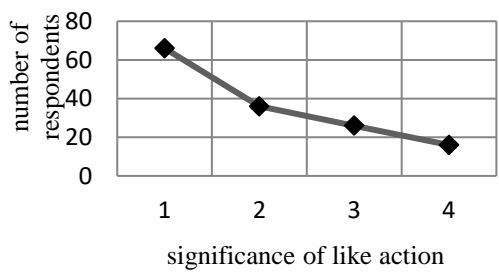

**Graph 2** Relation of likes on Facebook and closeness in real life

Finally, the participants had to choose one among six interaction parameters, which, in their opinion, represent the best strength of the friendship in the real life. Replies are shown on the graph 3.

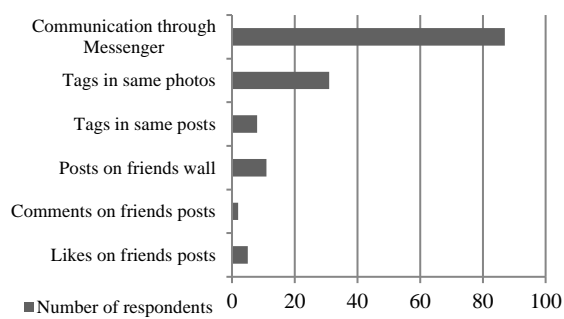

**Graph 3** Importance of parameters from user perspective

### 4.2 Used parameters and variables

According to previously described survey and the answers shown in graph 2, it was decided that all interaction parameters that involve *like* action won't be used. Such action is easily performed and users like posts from the best friends and acquaintances equally, depending on the content of a post.

Since 84% users believe that the number of mutual friends on Facebook and closeness in real life are not related, interaction parameter that carries such information also wasn't used.

Graph number 1 shows that users, who use the *Close Friends* list on Facebook, use it for their "real life" close friends. If more users applied this option, modeling would be facilitated, therefore, this interaction parameter was given a special importance.

Interaction parameter that carries the information about a number of messages exchanged is numerical parameter of the largest range and according to the user answers one of the most important parameters of interaction. Therefore, in the process development of the mathematical model it was given a special importance.

The first launch of the Linear Regression and the Random Forest algorithms has shown that interaction parameter that carries the information about the number of mutual photos published by user B impacts classification negatively, therefore that parameter won't be used in the proposed models.

Finally, 8 of 15 parameters available in the data set from previous research were used and complexity was reduced.

Numerical interaction parameters used in our model are described in the table 3. User A is an ego-user who has participated in the research and whose (subjective) assessment of the observed relationship is considered as ground truth, while user B is a person from A's friend list.

**Table 3** Used interaction parameters

| |
|---|
| User B *comment on a wall post* of user A |
| User B *message on the wall* of user A |
| Users A and B *tagged together in a post* |
| *Mutual photo* published by user A |
| *Mutual photo* published by a user that is not user A or B |
| User B comments on a photo of user A |
| *Messages* exchanged between user A and B |
| User A used *close friend* option for user B |

Parameter importance represented by the *t value* is obtained with the Linear Regression and given in the table 4.

**Table 4** Values of t value variable

| Parameter of interaction | *t value* |
|---|---|
| User B *comment on a wall post* of user A | 12,257 |
| User B *message on the wall* of user A | 3,397 |
| Users A and B *tagged together in a post* | 5,513 |
| *Mutual photo* published by user A | 2,763 |
| *Mutual photo* published by a user that is not user A or B | 30,157 |
| User B comments on a photo of user A | 0,789 |
| *Messages* exchanged between user A and B | 12,581 |
| User A has user B in *Close Friends* list | 22,837 |

Values of *MeanDecreaseAccuracy* for each interaction parameter are given in the table 5.

**Table 5** Values of MeanDecreaseAccuracy variable

| Parameter of interaction | MeanDecrease Accuracy |
|---|---|
| User B *comment on a wall post* of user A | 49,30 |
| User B *message on the wall* of user A | 59,53 |
| Users A and B *tagged together in a post* | 48,89 |
| *Mutual photo* published by user A | 48,58 |
| *Mutual photo* published by a user that is not user A or B | 45,95 |
| User B comments on a photo of user A | 37,47 |
| *Messages* exchanged between user A and B | 138,42 |
| User A has user B in *Close Friends* list | 102,65 |

Values of $p_i$ and values of variables used in their calculation are given in the table 6. Meanings of parameters *b*, *v* and *u* are described in sub-section 3.3.

Table 6 Parameters calculated according to user answers in a survey

| Parameter of interaction | b | v | u | $p_i$ |
|---|---|---|---|---|
| User B *comment on a wall post* of user A | 2 | 2 | 139 | **0,0072** |
| User B *message on the wall* of user A | 11 | 1 | 139 | **0,0791** |
| Users A and B *tagged together in a post* | 8 | 1 | 139 | **0,0576** |
| *Mutual photo* published by user A | 31 | 2 | 139 | **0,1115** |
| *Mutual photo* published by a user that is not user A or B | 31 | 2 | 139 | **0,1115** |
| User B comments on a photo of user A | 2 | 2 | 139 | **0,0072** |
| *Messages* exchanged between user A and B | 88 | 1 | 139 | **0,6330** |
| User A has user B in *Close Friends* list | | | | **1,0000** |

Model accuracy was verified with the previously described data that contains user answers about a better friend in a pair. In other words, the better friend determined by mathematical model (6) was compared with the better friend from user's statement. The accuracy was calculated for each of the parameters of interaction with coefficients $k_i$ from the interval [1, 100] and for greater random numbers above 100, while all other parameters were fixed as value 1. By using coefficient $k_i$ for parameter that carries information about the number of messages exchanged which is proved several times as the most important parameter, further improvement of the model is possible. To make it easier we will label this parameter as *inbox_chat*. Until $k_{inbox\_chat} = 7$ accuracy rises and for all numbers above 7 there is no change. Some greater random numbers above 100 were also tested and it was shown that the higher coefficient $k_{inbox\_chat}$ the accuracy of the model is falling.

Change in the accuracy with the use of experimentally determined coefficient that gives special importance to the parameter containing information about the number of exchanged messages between observed users is given in the table 7.

Table 7 Change of accuracy depending on coefficient $k_{inbox\_chat}$

| $k_{inbox\_chat}$ | Accuracy (12 598 pairs) |
|---|---|
| 1 | 82,32% |
| 2 | 83,12% |
| 3 | 83,55% |
| 4 | 83,63% |
| 5 | 83,74% |
| 6 | 83,75% |
| 7 | 83,80% |
| 10 | 83,80% |
| 50 | 83,80% |
| 100 | 83,80% |

For all the other parameters of interaction when $k_i$ is tested there is no further change in accuracy. For that reason, all other coefficients are set as 1. All coefficients $k_i$ are given in the table 8.

Table 8 Coefficients k for each interaction parameter

| *i*-th parameter of interaction | $k_i$ |
|---|---|
| User B *comment on a wall post* of user A | 1 |
| User B *message on the wall* of user A | 1 |
| Users A and B *tagged together in a post* | 1 |
| *Mutual photo* published by user A | 1 |
| *Mutual photo* published by a user that is not user A or B | 1 |
| User B comments on a photo of user A | 1 |
| *Messages* exchanged between user A and B | 7 |
| User A has user B in *Close Friends* list | 1 |

### 4.3 Verification of proposed models

All models' accuracy was verified with the previously described data that contains user answers about a better friend in a pair. With accuracy as a concept we demonstrate how well the mathematical model reproduces user answers. It is considered that the better friend in the offered pair is the one with the greater link weight calculated by using the mathematical model. The results were verified two methods. The first method was comparing prediction of the model and user answers on a complete data set. Second was comparing user answers with the model prediction without the pair of user friends for which predicted weights were equal because rationally such weights cannot be compared.

If *friendship_weight* for the first friend in the pair is greater than the second, observed friend is considered to be closer friend and vice versa.

After obtaining parameter importance with the Linear Regression algorithm, model's (2) accuracy is verified and the results are given in the table 9.

Table 9 Accuracy of model (2)

| Type of data set | Accuracy |
|---|---|
| Data without the predicted weights that are equal | 76,42% (9627 of 12 598) |
| Original data | 72,74% (9627 of 13 235) |

Subsequently parameter importance with the Random Forest algorithm is acquired, model's (3) accuracy is verified and the results are given in the table 10.

Table 10 Accuracy of model (3)

| Type of data set | Accuracy |
|---|---|
| Data without the predicted weights that are equal | 78,80% (9927 of 12 598) |
| Original data | 75,00% (9927 of 13 235) |

Compared to the model (2), the model (3) gives approximately 2,30% better results and it's decided that the parameter importance can be represented by value of the *MeanDecreaseAccuracy* variable. If in addition to this variable we use proposed $p_i$ values for each interaction parameter, the model (5) has accuracy shown in the table 11, which is approximately 3,4 % greater than the accuracy of the model (3).

Table 11 Accuracy of model (5)

| Type of data set | Accuracy |
|---|---|
| Data without the predicted weights that are equal | 82,32% (10 370 of 12 598) |
| Original data | 78,35% (10 370 of 13 235) |

The final model is given with (6) and its accuracy is shown in the table 12. Compared to the model (5) it gives approximately 1,5% better accuracy if verified with the same data.

Table 12 Accuracy of the final model (6)

| Type of data set | Accuracy |
|---|---|
| Data without predicted weights that are equal | 83,80% (10 557 of 12 598) |
| Original data | 79,77% (10 557 of 13 235) |

Accuracy of the proposed final model was also compared to the similar mathematical model [8], which was verified with the same data set. This model uses all of the 15 interaction parameters from the original data set, unlike the model from this paper, which uses a reduced set of 8 parameters. Before verifying, all data were normalized equally as previously described in the section 3.2. In comparison with the model from previous research [8], which employs all available interaction parameters, our final model (6) gives better results by 1,07%.

## 5 Discussion

The research results of the proposed final model (6) described in this paper over an available data set gives a satisfactory result. It is possible to determine the intensity of friendship between OSN users based on their interaction with a high accuracy (83,80%).

Better insight into interaction parameters from the users' perspective was obtained through a survey. Survey discovered a big difference in the predictive strength between different types of interaction. For example, a greater number of likes and mutual friends doesn't imply greater closeness in real life, therefore those parameters weren't used. Also, it was noticed that number of exchanged messages and close friends option have a great predictive strength so we decided to focus more on these parameters.

Predictive mathematical model was built using supervised machine learning algorithms: Linear Regression and Random Forest. These algorithms were used to obtain the importance of each interaction parameter observed. Random Forest algorithm gave better results than Linear Regression. It was shown that accuracy rises with the application of experimentally determined coefficient as a parameter that carries information about the number of exchanged messages, and by utilizing distribution parameters calculated according to users' opinion instead of the model which solely uses variable importance obtained with the Random Forest algorithm.

In comparison with the previous research which uses all available interaction parameters, our final model (6) gives 1,07% better accuracy observed on the same data set. Based on that, it can be concluded that it is possible to use a smaller set of interaction parameters and less complex mathematical model to get better results.

## 6 Conclusion and future work

Online social networks are one of the most popular means of modern communication. Among the best known is Facebook which has an average of one billion daily active users. It is of a great value to identify more trusted, close and in general like-minded friends of an observed user in order to perform targeted recommendations of products and services or to expose a network of criminals, terrorists or fraudsters. Also, it can be applied in education process for detecting influential students.

The aim of this research was to design and verify the mathematical model using anonymized data from Facebook collected in previous research to determine the intensity of friendship between online social network users based on their interaction.

A survey was conducted through a questionnaire that was answered by 144 people who mostly use Facebook daily in order to obtain a better insight into specific interaction parameter importance from the perspective of users. The goal of the survey was to find out which parameters to use and how to use them when designing a mathematical model. It was concluded that one of the most valuable parameters from the user perspective is a number of private messages exchanged. According to survey responses, interesting information opposite to some previous research results was discovered – usage of the like action depends on the content of the post regardless of the "real life" relationship status. Therefore, it was decided that such action, including the number of mutual friends, will not be taken into consideration in the process of modelling because respondents as originators of interaction considered it not to be associated with the intimacy in real life.

The model was built using Random Forest algorithm to define the importance of all parameters which carry information about the amount of interaction between users and their friends on Facebook. According to the research and user answers shown in graph 3, distribution of interaction parameters was calculated. Also, coefficient value for the most important parameter of interaction, the number of exchanged messages, was determined experimentally.

The accuracy of the final proposed model, verified by using answers about the better friend in the offered pair, is (83,80%). It confirms the hypothesis of the research, which states that it is possible to determine friendship intensity based on interaction analysis between users on OSN. Also, it is important to emphasize that our model has lower complexity, but higher accuracy in comparison with the model from previous research.

Future work should include experimenting with and testing of other supervised machine learning algorithms in order to rise the accuracy of the proposed model by determining the importance of interaction parameters. The survey conducted in this research could include a larger number of participants and enhanced distribution of interaction parameters calculated.

### Acknowledgements

The work has been fully supported by Croatian Science Foundation under the project (UIP-2014-09-2051 eduMINE – Leveraging data mining methods and open technologies for enhancement of the e-learning infrastructure).

**Authors' addresses**

*Sanja Krakan, M.Sc. in Information and Communication Technology*
University of Zagreb, Faculty of Electrical Engineering and Computing
Unska 3, 10000 Zagreb, Croatia
sanja.krakan@outlook.com
+39 366 5476 307, +385 91 3003 403

*Luka Humski, PhD student* (Corresponding Author)
University of Zagreb, Faculty of Electrical Engineering and Computing
Unska 3, 10000 Zagreb, Croatia
+385 1 6129 763, +385 95 803 95 69, luka.humski@fer.hr

*Zoran Skočir, PhD, Full Professor*
University of Zagreb, Faculty of Electrical Engineering and Computing
Unska 3, 10000 Zagreb, Croatia
zoran.skocir@fer.hr